\documentclass{statsoc}

\usepackage[english,activeacute]{babel}
\usepackage[a4paper]{geometry}
\usepackage{graphicx}
\usepackage[textwidth=8em,textsize=small]{todonotes}
\usepackage{amsmath}
\usepackage{amssymb}
\usepackage{natbib}
\usepackage{url}
\usepackage{enumerate}
\usepackage{multicol}
\usepackage{multirow}
\usepackage[T1]{fontenc}
\usepackage{float}
\usepackage{subfigure}
\usepackage{rotating}
\usepackage{ulem}

\restylefloat{table}

\def\exp#1{{\rm exp}{#1}}
\def\frac#1#2{{{#1}\over{#2}}}

\def\le{\left}
\def\ri{\right}

\newcommand\simiid{\mathrel{\overset{\makebox[0pt]{\mbox{\normalfont\tiny\sffamily iid}}}{\sim}}}
\newcommand\simind{\mathrel{\overset{\makebox[0pt]{\mbox{\normalfont\tiny\sffamily ind}}}{\sim}}}

\newcommand{\ind}[1]{\mathbb{I}\left\{ #1 \right\}}
\newcommand{\pr}[1]{\mathbb{P}\text{r}\left[#1\right]}
\newcommand{\expec}[1]{\mathbb{E}\left[#1\right]}

\newcommand{\sd}[1]{\text{SD}\left[#1\right]}

\newcommand{\coefvar}[1]{\text{CV}\left[#1\right]}

\newcommand{\ex}[1]{\exp{ \left\{ #1 \right\}}}
\newcommand{\quo}[1]{\textquotedblleft#1\textquotedblright}

\newcommand{\TP}{\text{TP}}

\newcommand{\FP}{\text{FP}}
\newcommand{\FN}{\text{FN}}

\def\I{\mathbf{I}}

\def\P{\mathbf{P}}\def\pv{\boldsymbol{p}}

\def\uv{\boldsymbol{u}}

\def\wv{\boldsymbol{w}}
\def\X{\mathbf{X}}\def\x{\mathbf{x}}
\def\Y{\mathbf{Y}}\def\y{\mathbf{y}}

\def\al{\alpha}\def\alv{\boldsymbol{\alpha}}
\def\be{\beta}
\def\ga{\gamma}

\def\del{\delta}

\def\ze{\zeta}\def\zev{\boldsymbol{\zeta}}

\def\vte{\vartheta}\def\vtev{\boldsymbol{\vartheta}}

\def\si{\sigma}
\def\sig{\sigma}

\def\ome{\omega}

\def\xiv{\boldsymbol{\xi}}

\def\piv{\boldsymbol{\pi}}

\def\UPS{\mathbf{\Upsilon}}

\def\Cat{\small{\mathsf{Cat}}}
\def\Dir{\small{\mathsf{Dir}}}

\def\Ber{\small{\mathsf{Ber}}}

\def\Nor{\small{\mathsf{N}}}

\def\Bet{\small{\mathsf{Beta}}}

\def\IGamd{\small{\mathsf{IGam}}}

\def\data{\text{data}}

\def\rest{\text{rest}}

\def\rest{\text{rest}}

\def\yiij{y_{i,i',j}}

\def\xivhat{\hat{\boldsymbol{\xi}}}

\def\zerov{\boldsymbol{0}}
\def\onev{\boldsymbol{1}}

\title[A Record Linkage Model Incorporating Relational Data]{A Record Linkage Model Incorporating Relational Data}
\author[Juan Sosa]{Juan Sosa}
\address{Universidad Externado de Colombia}
\email{juan.sosa@uexternado.edu.co}
\author{Abel Rodr\'iguez}
\address{University of California, Santa Cruz}
\email{abel@soe.ucsc.com}

\begin{document}

\begin{abstract}

In this paper we introduce a novel Bayesian approach for linking multiple social networks in order to discover the same real world person having different accounts across networks.  In particular, we develop a latent model that allow us to jointly characterize the network and linkage structures relying in both relational and profile data.  In contrast to other existing approaches in the machine learning literature, our Bayesian implementation naturally provides uncertainty quantification via posterior probabilities for the linkage structure itself or any function of it.  Our findings clearly suggest that our methodology can produce accurate point estimates of the linkage structure even in the absence of profile information, and also, in an identity resolution setting, our results confirm that including relational data into the matching process improves the linkage accuracy.  We illustrate our methodology using real data from popular social networks such as \texttt{Twitter}, \texttt{Facebook}, and \texttt{YouTube}.

\vspace{6pt}

\hspace{6pt} \textit{Keywords.}  Entity Resolution; Network data; Latent space models; Social network analysis; Record Linkage.
\end{abstract}

\section{Introduction}



Online social networks (OSNs) have an enormous impact in several human facets. Thanks to the quick development of mobile devices, people tend to split their social activities into several OSNs: They look for family and friends in \verb"Facebook", express their points of views in \verb"Tweeter", and share their photographs in \verb"Instagram", just to give a few examples. Such diversity suggests that studies attempting to understand online social dynamics by including just one single OSN will end up having a strong bias, and therefore, they will typically be incorrect.


Individuals involved in two or more OSNs may employ different profile names with some discrepancies about their profile information.  In addition, some accounts may have similar profile information, but they still correspond to different users. Hence, OSNs integration is an active area of research.





Merging OSNs is of great importance for three major reasons. First, creating more comprehensive social databases for entrepreneurs and scientists. Unified data sources prevent double-counting in research studies involving OSNs and provide invaluable information to test scientific hypothesis. Second, developing automatic contacts merging in mobile devices. Often advanced users provide full access to several OSNs in order to somehow integrate the data streams from their contacts. And third, re-identifying (de-anonymizing) social networks to analyze privacy matters and online security issues.


Databases (files) often contain corrupted data with duplicate entries across and within each database. Merging databases and/or removing duplicated records by grouping records pertaining to the same entities lacking unique identifiers is known as record linkage (RL) or entity resolution (ER). Such problem can be viewed as a type of  ``microclustering'', with few observations per cluster and a very large number of clusters. This task has applications in all sorts of settings, including public health, human rights, fraud detection, and national security. In this manuscript, we focus our attention on a meaningful RL problem: Identifying accounts across OSNs owned by the same user.


Several methods have been studied from the machine learning/computer science perspective for merging OSNs. \citet{vosecky-2009} and \citet{veldman-2009} combine profile features with weights to determine matches through a threshold-based approach. \citet{malhotra-2012} and \citet{nunes-2012} adopt supervised classification to decide on matching. Such methods mainly rely on pairwise profile comparisons, which are computationally very expensive; as a matter of fact, many actors are so dissimilar that there is no need to compare them. In order to reduce computational burden, \citet{zhang-2014} develop a classifier setting up soft constraints to form overlapping clusters to prune unnecessary pairwise comparisons. Beyond profile comparison, \citet{narayanan-2009} introduce a de-anonymization algorithm purely based on network topology, by exploiting the fact that users often have similar social connections in different OSNs. \citet{bartunov-2012} attack the problem employing conditional random fields that account for both profile attributes and social linkage. Their findings confirm that including graph information improves performance by re-identifying users with similar relationship structures. Concurrently, also considering both sources of information, \cite{buccafurri-2012} propose a neighborhood similarity approach to address the problem of multiple social network integration. More recently, \citet{zhang-2015} develop a learning algorithm based on dual decomposition for the same purpose. \citet{shu-2017} review key achievements of user identity linkage across OSNs.


Within the Bayesian paradigm to date, there has been significant advances in clustering and latent variable modeling, but to the best of our knowledge there are no references dealing with the problem of merging OSNs. The seminal work of \citet{fellegi-1969} is the classical reference for an unsupervised approach to find links between co-referent records. \citet{belin-1995}, \citet{fienberg-1997}, \citet{larsen-2001} and \citet{tancredi-2011} develop valuable methods to merge two files using samples of multivariate categorical variables. However, these techniques do not easily generalize to either multiple files or duplicate detection. \citet{gutman-2013} devise a general procedure that jointly models both the association between variables and the linkage structure. \citet{liseo-2013} offer a nice review of early Bayesian methodologies for performing RL and making inference using matched units.


More recently, \cite{steorts-2014-SMERED}, \cite{steorts-2015-empirical}, and \cite{steorts-2015-graphical}, extending ideas from \citet{domingos-2004}, propose an unsupervised method for linking records across arbitrarily many files, while simultaneously detecting duplicate records within files. Their key innovation consists in representing the pattern of links between records as a bipartite graph, in which records are directly linked to latent true individuals, and only indirectly linked to other records.



Even though these models are generative and embedded in the Bayesian paradigm, they rely on field-based records (unimodal data). In this manuscript we propose to extend the bipartite graph approach of \cite{steorts-2015-graphical} by incorporating relational information (bimodal data) in the context of OSN integration. Our proposal can be employed whether network information is available or not, and it is able to handle multiple files simultaneously.


The remainder of this manuscript is organized as follows: Section \ref{sec_motivating_examples} motivates our approach in two different settings, namely, re-identification (de-anonymization) and identity resolution. Section \ref{sec_a_RL_model_using_relational_data} introduces a novel approach for RL handling both profile and network data; there, we discuss in detail every aspect of the model including prior specification and computation. Section \ref{sec_posterior_linkage} presents some approaches to posterior linkage estimation and performance assessment. Sections \ref{sec_buccafurri_dataset} and \ref{sec_bartunov_dataset} revisit our motivating examples introduced in Section \ref{sec_motivating_examples} and present analyzes based on our proposed model. Finally, some concluding remarks are provided in Section \ref{sec_extensions}.


\section{Motivating examples}\label{sec_motivating_examples}

Here we present two problems to motivate our methodological developments in Section \ref{sec_a_RL_model_using_relational_data}. Even though the examples below arise in different contexts, they easily fit in a RL framework in which no more than two records correspond to the same entity, an also, de-duplication is not required.

\subsection{Re-identification}

\citet{bartunov-2012} considered a local RL setting where the goal is to match (re-identify) anonymized profiles across the contacts of a particular seed user. This task, known as re-identification or de-anonymization, includes several practical applications such as contact merging in mobile devices.

\begin{figure}[!t]
	\centering
	\includegraphics[scale=.55,]{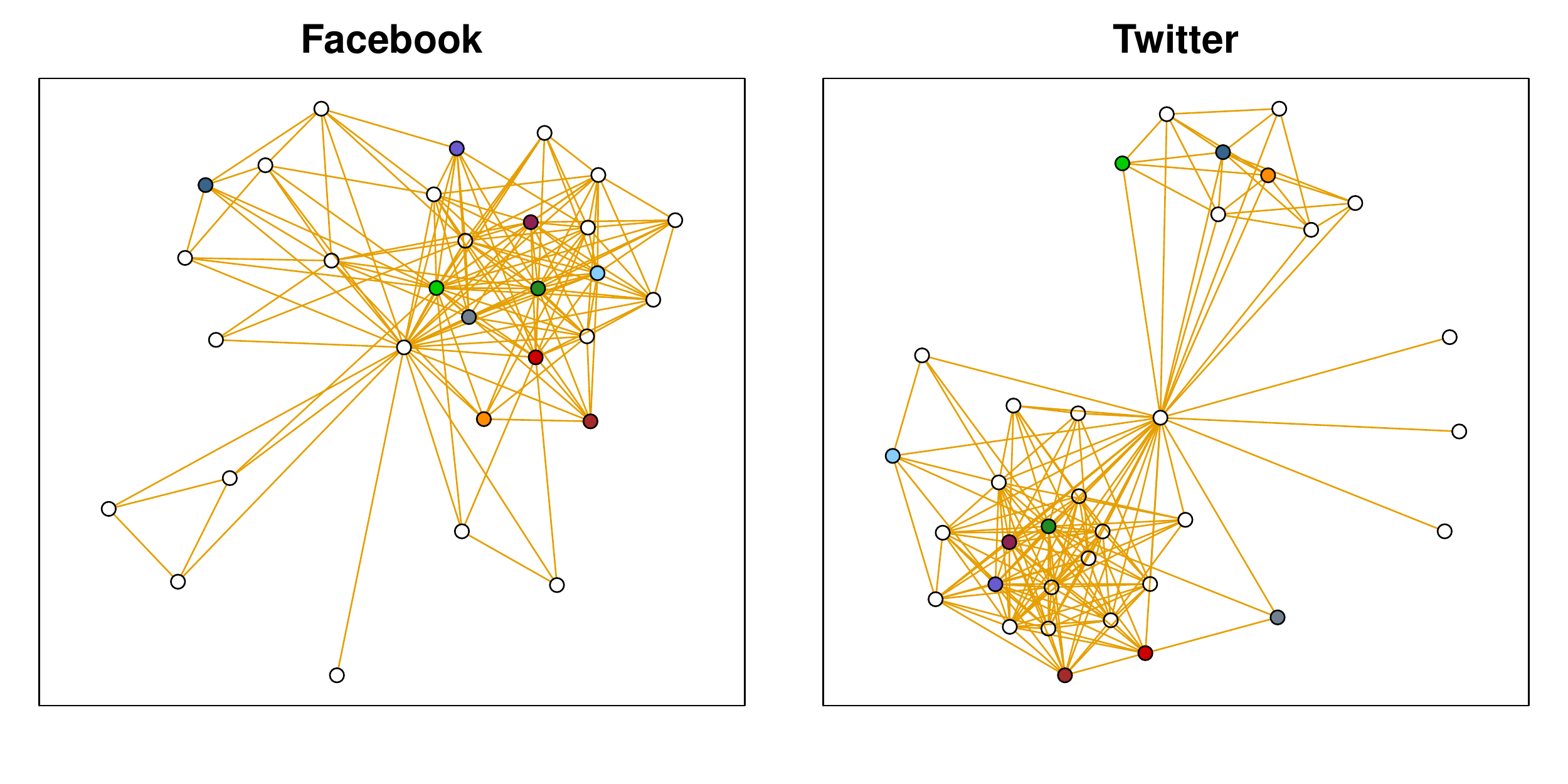}
	\caption{\footnotesize{Facebook and Twitter ego-networks for a particular user (seed node ID 32526). Matches are represented with same color vertices.}}
	\label{fig_nets_32526}
\end{figure}

Specifically, for 17 users, a first order snowball sample was taken from \verb"Facebook" and \verb"Twitter", collecting the first order mutual contacts along with their relations (first neighborhood); the authors share an anonymized version of the data at \url{http://modis.ispras.ru/uir/}. Thus, the full dataset consists of 17 anonymized pairs of ego-networks (one pair for each seed user) with manually mapped right projections (ground truth). As a final remark, the authors consider only mutual following in \verb"Twitter" in order to simulate friendship relationships as in \verb"Facebook". Hence, for every seed user, we have two undirected, binary networks with no profile information (all data are anonymized).


\begin{table}[!b]
	\centering
	\begin{tabular}{l|cccc|cccc}
		\hline
		& \multicolumn{4}{c|}{Facebook} & \multicolumn{4}{c}{Twitter}  \\  \hline
		Statistic               & Mean  & 25\%  & 50\%  & 75\%  & Mean  & 25\%  & 50\%  & 75\%  \\ \hline
		Density                 & 0.21  & 0.15  & 0.18  & 0.27  & 0.17  & 0.11  & 0.13  & 0.21  \\
		Clustering coefficient  & 0.30  & 0.14  & 0.29  & 0.40  & 0.43  & 0.36  & 0.40  & 0.52  \\
		Assortativity index     & -0.53 & -0.63 & -0.52 & -0.45 & -0.29 & -0.42 & -0.24 & -0.21 \\ \hline
	\end{tabular}
	\caption{\label{tab_data_stats_uir}\footnotesize{Bartunov's data summary statistics.}}
\end{table}

Figure \ref{fig_nets_32526} displays one pair of these networks for a particular user (seed node ID 32526). Summary statistics for these data are provided in Table \ref{tab_data_stats_uir}. Note that the networks are quite dense and exhibit a high level of transitivity; the negative assortativity values shown there might be explained by the egocentric nature of the networks. In addition, the total number of matches (also known as anchor nodes in this context) is 152; the ratio of true number of users to total number of nodes ranges from 0.0465 to 0.1915.

\subsection{Identity resolution}

\begin{figure}[h!]
	\centering
	\includegraphics[scale=.55,]{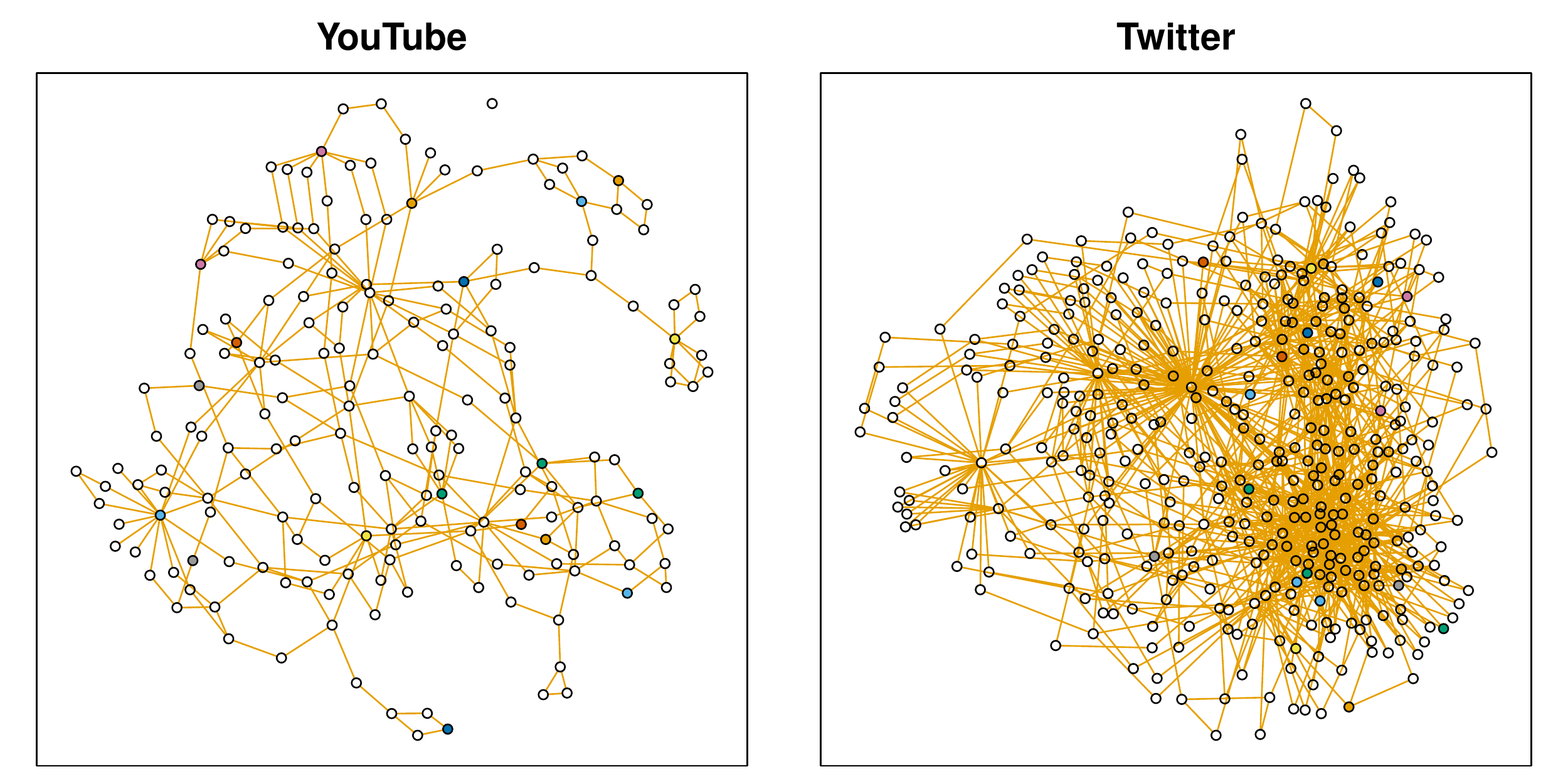}
	\caption{\footnotesize{YouTube and Twitter graphs. Matches are represented with same color vertices.}}
	\label{fig_nets_pkdd}
\end{figure}

\cite{buccafurri-2012} faced the problem of determining profiles owned by the same individual across different OSNs, based on both network topology and publicly available personal information. Matching such profiles allows to create a ``global'' profile that gives a complete view of an user's information. To this end, the authors collected data from two popular OSNs, namely, \verb"YouTube" and \verb"Twitter", which is available at \url{http://www.ursino.unirc.it/pkdd-12.html}. This dataset consists of two undirected, binary networks (only mutual following was considered, see Figure \ref{fig_nets_pkdd}) along with the username of each actor. The ground truth (correct matches) is also available.

For our illustrative purposes, we consider a subset of Buccafurri's data. Such subset contains 184 \verb"YouTube" users and 420 \verb"Twitter" users, 19 of which (3.2\%) correspond to the same individual; our goal is to identify as many of those matches as possible relying on their username, which might be distorted, and their connections. For instance, we expect to match \verb"banditooisclutch" in \verb"YouTube" and \verb"negritobandito" in \verb"Twitter" (since the ground truth tell us so), helped by the corresponding relational patterns (they are well connected users, degree 5 and 11, respectively), even though their usernames are somewhat dissimilar.

Summary statistics for these data are provided in Table \ref{tab_data_stats_pkdd}. Even though both networks seem to be alike in terms of density, \verb"Twitter"'s graph shows a higher degree of transitivity. The assortative values indicate some negative correlation between nodes of different degree, which to some extent may favor matching.

\begin{table}[!h]
	\centering
	\begin{tabular}{l|cc} \hline
		Statistic               & \verb"YouTube" & \verb"Twitter" \\  \hline
		Density                 & 0.0162  & 0.0122       \\
		Clustering coefficient  & 0.007   & 0.0366       \\
		Assortativity index     & -0.2405 & -0.1728      \\ \hline
	\end{tabular}
	\caption{\label{tab_data_stats_pkdd}\footnotesize{Buccafurri's data summary statistics.}}
\end{table}

\section[A record linkage model]{A record linkage model incorporating relational data}\label{sec_a_RL_model_using_relational_data}

We begin by specifying some notation and the key idea behind our approach. Then, we present all the details about the model and the prior specification. Finally, remarks about hyperparameter elicitation and computation are provided.

\subsection{Notation}\label{sec_notation_RL2}




We consider a collection of $J\geq2$ undirected, binary networks with adjacency matrices $\Y_j=[\yiij]$, such that $\yiij = 1$ if there is a link between actors $i$ and $i'$ in network $j$, and $\yiij=0$ otherwise, $1\leq i < i'\leq I_j$. On the other hand, let $\pv_{i,j}=(p_{i,j,1},\ldots,p_{i,j,L})$ be the profile (attribute) data associated with the $i$-th actor in network $j$, and let $\P_{j}=[p_{i,j,\ell}]$ be the corresponding $I_j\times L$ array for every $j$. For simplicity, we assume that every profile contains $L$ fields in common, field $\ell$ having $M_\ell$ levels. Profile data of this sort may be considered as either categorical or string-valued. Let us say, for instance, that data about gender, state of residency, and race regarding $I_j$ actors in network $j$ are available; in this scenario, $\pv_{i,j}$ is a categorical vector with dimension $L=3$ whose entries have $M_1=2$ (male and female), $M_2=50$ (as of 2017, there are 50 states in the United States), and $M_3=6$ (White, Black or African-American, American Indian or Alaska Native, Asian, Native Hawaiian or Other Pacific Islander, and some other race) levels, respectively. Hence, we can think of records as $L + (I_j - 1)$ dimensional vectors storing profile information ($L$ fields) as well as link information ($I_j-1$ binary fields), while the $j$-th file, in turn, is composed of $I_j$ records.


Now, let $\piv_n=(\pi_{n,1},\ldots,\pi_{n,L})$ be the vector of ``true'' attribute values for the $n$-th latent individual, $n=1,\ldots,N$, where $N$ is the total number of latent individuals in the population ($N$ could be as small as $1$ if every record in every file refers to the same entity or as large as $I = \sum_j I_j$ if files do not share records at all). Hence, $\piv=[\pi_{n,\ell}]$ is an unobserved $N\times L$ attribute matrix whose $n$-th row stores the profile data associated with the $n$-th latent individual. Next, we define the linkage structure $\xiv=(\xiv_{1},\ldots,\xiv_{J})$, where $\xiv_j=(\xi_{1,j},\ldots,\xi_{I_j,j})$. Here, $\xi_{i,j}$ is an integer from 1 to $N$ indicating which latent individual the $i$-th record in file $j$ refers to, which means that $\pv_{i,j}$ is a possibly-distorted measurement of $\piv_{\xi_{i,j}}$. Such structure unequivocally defines a partition $\mathcal{C}_{\xiv}$ on $\{1,\ldots,I\}$. To see this, notice that by definition, two records $(i_1,j_1)$ and $(i_2,j_2)$  correspond to the same individual if and only if $\xi_{i_1,j_1}=\xi_{i_2,j_2}$. Therefore, $\mathcal{C}_{\xiv}$ is nothing more than a set composed of $N$ disjoint non-empty subsets $\{C_1,\ldots,C_N\}$ such that $\cup_n C_n = \{1,\ldots,I\}$ where each $C_n$ is defined as the set of all records pointing to latent individual $n$. Hence, the total number of latent individuals $N=N(\xiv)$ is a function of the linkage structure; specifically, $N=\max\{\xi_{i,j}\}$, since for simplicity we have labeled the cluster assignments with consecutive integers from 1 to $N$. Lastly, $w_{i,j,\ell}$ is a binary variable defined as 1 or 0 according to whether or not a particular field $\ell$ is distorted in $\pv_{i,j}$, i.e.,
$$
w_{i,j,\ell} = \left\{
\begin{array}{ll}
1, & \hbox{ $p_{i,j,\ell} \neq \pi_{\xi_{i,j},\ell}$;} \\
0, & \hbox{ $p_{i,j,\ell} = \pi_{\xi_{i,j},\ell}$.}
\end{array}
\right.
$$
Then, each $\wv_j=[w_{i,j,\ell}]$ is a $I_j\times L$ binary matrix containing the (unobserved) distortion indicators of the profile data in file $j$.

For example, suppose that the (latent) population has $N=4$ members and they are listed as before by gender, state and race. To illustrate, let the latent population matrix $\piv$ be
$$
{\small
	\piv = \begin{bmatrix}
	\text{F} & \text{CA} & \text{White} \\
	\text{F} & \text{NY} & \text{Black} \\
	\text{M} & \text{MI} & \text{Asian} \\
	\text{M} & \text{CA} & \text{White} \\
	\end{bmatrix}.
}
$$
We also consider $J=3$ files with $I_1=4$, $I_2=5$, and $I_3=4$ users, whose (observed) profiles might be
$$
{\small
	\P_1 = \begin{bmatrix}
	\text{F} & \text{\textbf{CA}} & \text{Black} \\
	\text{M} & \text{\textbf{RI}} & \text{Asian} \\
	\text{M} & \text{CA} & \text{White} \\
	\text{F} & \text{\textbf{MA}} & \text{White} \\
	\end{bmatrix},\,\,
	\P_2 = \begin{bmatrix}
	\text{F} & \text{\textbf{LA}} & \text{White} \\
	\text{M} & \text{MI} & \text{Asian} \\
	\text{F} & \text{\textbf{NJ}} & \text{Black} \\
	\text{M} & \text{CA} & \text{White} \\
	\text{M} & \text{CA} & \text{White} \\
	\end{bmatrix},\,\,
	\P_3 = \begin{bmatrix}
	\text{M} & \text{\textbf{IA}} & \text{White} \\
	\text{F} & \text{\textbf{PA}} & \text{White} \\
	\text{F} & \text{\textbf{NV}} & \text{Black} \\
	\text{M} & \text{MI} & \text{Asian} \\
	\end{bmatrix}.
}
$$
Here, for the sake of keeping the illustration simple, only state is distorted. Thus, comparing the observed profiles $\P_1$, $\P_2$ and $\P_3$ to the latent population $\piv$, the corresponding linkage structure and distortions indicators are $\xiv_1 =(2, 3, 4, 1)$, $\xiv_2=(1,3,2,4,4)$, $\xiv_3=(4,1,2,3)$,
$$
{\small
	\wv_1 = \begin{bmatrix}
	0 & 1 & 0 \\
	0 & 1 & 0 \\
	0 & 0 & 0 \\
	0 & 1 & 0 \\
	\end{bmatrix},\,\,
	\wv_2 = \begin{bmatrix}
	0 & 1 & 0 \\
	0 & 0 & 0 \\
	0 & 1 & 0 \\
	0 & 0 & 0 \\
	0 & 0 & 0 \\
	\end{bmatrix},\,\,
	\wv_3 = \begin{bmatrix}
	0 & 1 & 0 \\
	0 & 1 & 0 \\
	0 & 1 & 0 \\
	0 & 0 & 0 \\
	\end{bmatrix}.
}
$$
Every entry of each $\xiv_j$ with a value of 2 means that the corresponding profile in $\P_j$ refers to the latent individual with attributes \quo{F}, \quo{NY} and \quo{Black}. The state of this individual has been distorted in all three files as can be seen from every $\wv_j$. We also see other profiles distorted in each $\wv_j$.


\begin{figure}[h!]
	\centering
	\includegraphics[scale=.38,]{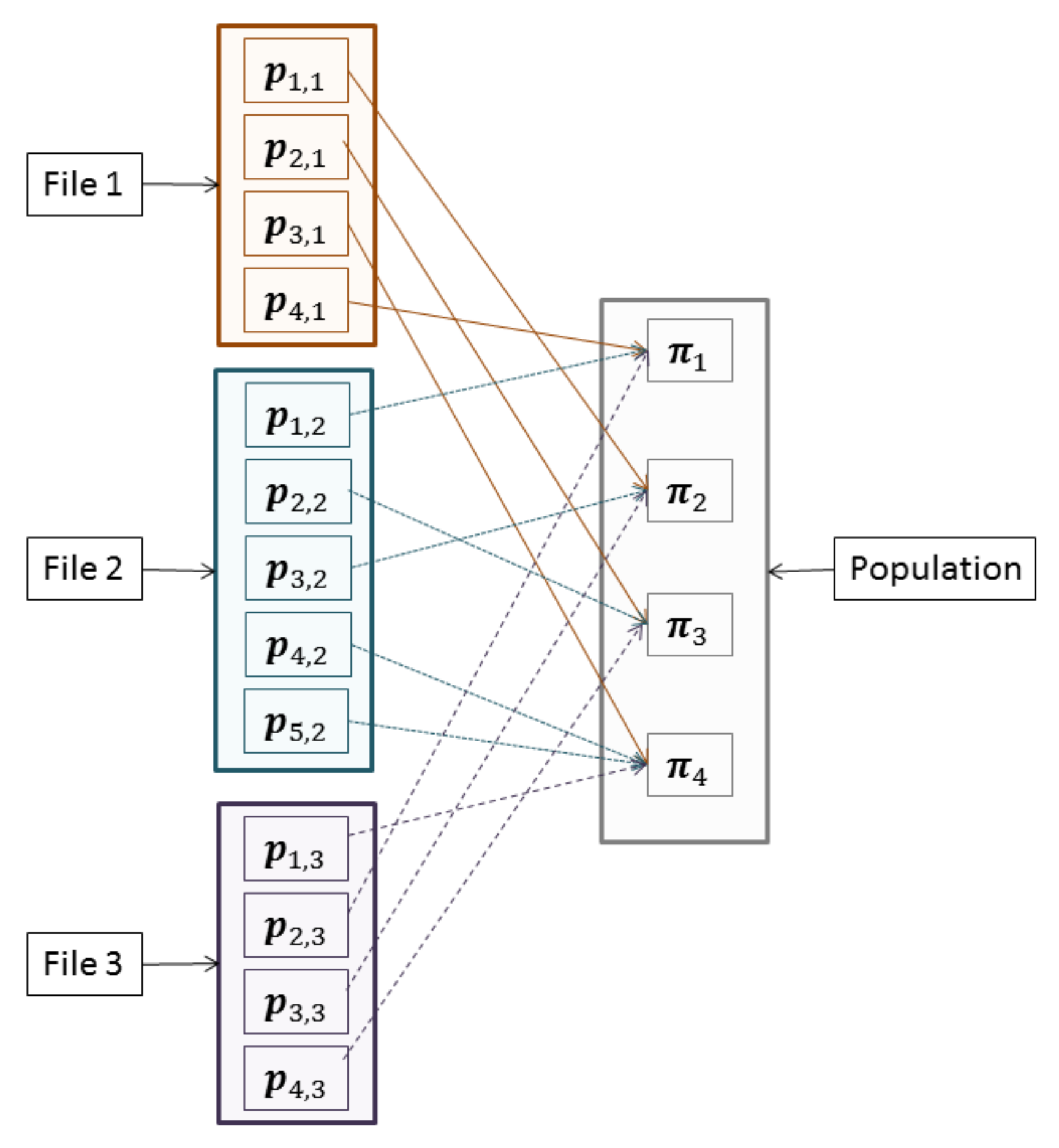}
	\caption{\footnotesize{Illustration of profiles $\pv_{i,j}$, latent true profiles $\piv_{n}$, and linkage structure (edges) $\xiv$.}}
	\label{fig_bipartie}
\end{figure}

Figure \ref{fig_bipartie} shows the linkage structure $\xiv$ as a bipartite graph in which each edge links a profile to a latent individual. For instance, this figure shows that $\pv_{4,1}$, $\pv_{1,2}$, $\pv_{2,3}$ correspond to the same individual. This toy example makes clear that linking records to a hypothesized latent entity is at its core a clustering problem where the main goal is to make inferences about the cluster assignments $\xiv$. In contrast to other clustering tasks, we aim to develop an approach that let the number of records in each cluster be small, even for large datasets; such property is signature in RL and de-duplication applications.

In order to incorporate the network information, we will assume that every unique individual $n$ in the population has associated with it a latent position $\uv_n=(u_{n,1},\ldots,u_{n,K})$ of unobserved characteristics, which somehow get \quo{distorted} into the observed links. Notice that $\uv_n$ plays the same role of $\pv_n$, the same way that $\Y_j$ plays the same role of $\P_j$. Such mechanism is the key to relate the linkage structure with the profile and network data.

\subsection{Model formulation}\label{sec_model_formulation_RL2}

Here we present an approach that combines both profile data $\P=[p_{i,j,\ell}]$ and network data $\Y=[y_{i,i',j}]$ in order to uncover multiple records across different files corresponding to the same individual. By assuming that these two sources of information can be modeled independently given the linkage structure $\xiv$, our approach handles them separately just having the cluster assignments in common.

As before, given the cluster assignments $\xi_{i,j}$ and the latent characteristics $u_{n,k}$, we first model the network data $y_{i,i',j}$ conditionally independent using a latent distance model \citep{hoff-2002},
\begin{equation}\label{eq_model_RL_part_1}
y_{i,i',j}\mid \beta_j, \uv_{\xi_{i,j}},\uv_{\xi_{i',j}} \simind \Ber\le(g^{-1}\le(\be_j - \|\uv_{\xi_{i,j}} - \uv_{\xi_{i',j}}\|\ri)\ri),
\end{equation}
where $g(\cdot)$ is a (known) link function and $\be_j$ is a global intercept associated with file $j$.



Regarding the profile data, we model each field depending on whether it is distorted or not.  If $p_{i,j,\ell}$ is not distorted, i.e.,  $w_{i,j,\ell}=0$, we simply keep that particular field intact by giving it a point mass distribution at the true value.
On the other hand, if a distortion is present, i.e., $w_{i,j,\ell}=1$, we attempt to recover the real attribute by placing a categorical (multinomial) distribution over all the categories of that particular field. In summary, assuming that the profile data $p_{i,j,\ell}$ are conditional independent given the cluster assignments $\xi_{i,j}$ and the true population attributes $\pi_{n,\ell}$, we have that:
\begin{equation}\label{eq_model_RL_part_2}
p_{i,j,\ell}\mid\pi_{\xi_{i,j},\ell},w_{i,j,\ell},\zev_{\pi_{\xi_{i,j},\ell},\ell}\simind \left\{
\begin{array}{ll}
\del_{\pi_{\xi_{i,j},\ell}},        & \hbox{$w_{i,j,\ell}=0$;} \\
\Cat\le( \zev_{\pi_{\xi_{i,j},\ell},\ell} \ri), & \hbox{$w_{i,j,\ell}=1$,} \\
\end{array}
\right.
\end{equation}
where $\delta_a$ is the distribution of a point mass at $a$ and $\zev_{\pi_{\xi_{i,j},\ell},\ell}$ is an $M_\ell$-dimensional vector of multinomial probabilities that may or may not depend on $\pi_{\xi_{i,j},\ell}$.


Early approaches considered a probability vector indexed only by $\ell$, say $\vtev_\ell$, which is often taken as the prior probability for the true field \citep{steorts-2014-SMERED}. However, this ignores the fact that the kind of distortion we observe might depend on the value of the true field (e.g., finding state field distorted to CO might be more likely if the true value of the field is CA than WY). When $M_\ell$ is small we can model the corresponding profile values without imposing constraints.  On the contrary, if field $\ell$ has too many categories, a different approach is needed in order to reduce the number of parameters. For instance, in the case of string-valued fields, a common approach consists in letting the probability that a distorted value takes the value $s$ be
\begin{equation}\label{eq_emipirical_distr_string_field}
\ze_{\pi_{\xi_{i,j},\ell},\ell,s} \propto \gamma_\ell(s)\,\ex{-\lambda\,d\le(s,\pi_{\xi_{i,j},\ell}\ri)},
\end{equation}
where $\gamma_\ell(s)$ is equal to the empirical frequency of $s$ in field $\ell$, $\lambda > 0$ is a known real value, and $d(\cdot,\cdot)$ is some string distance. The probability distribution \eqref{eq_emipirical_distr_string_field} was originally considered as a distortion model
in \cite{steorts-2015-empirical}, proving to be successful in their applications.


The next stage of the model complements the formulation given in (\ref{eq_model_RL_part_1}) and (\ref{eq_model_RL_part_2}). First, we consider the latent effects $u_{n,k}$ as mutually independent and let $\uv_n\mid\sig^2\simiid \Nor(\zerov,\sig^2\,\I_K)$. Next, we let the true profile attributes to follow a categorical distribution by placing $\pi_{n,\ell}\mid\vtev_{\ell} \simind \Cat(\vtev_{\ell})$. Then, given that the distortion indicators $w_{i,j,\ell}$ are binary variables, we simply let $w_{i,j,\ell}\mid\psi_\ell\simind \Ber(\psi_\ell)$, where for simplicity each distortion probability $\psi_\ell$ depends only on the field.


Finally, we complete the model by specifying the independent priors: $\beta_j \sim \Nor(0,\ome_j^2)$, $\sig^2\sim\IGamd(a_\sig,b_\sig)$, $\vtev_{\ell} \simind \Dir(\alv_{\ell})$, $\psi_\ell  \simind \Bet(a_\ell,b_\ell)$, and $\xiv\sim p(\xiv)$. Hence, our model constitutes a full hierarchical Bayesian framework for discovering duplicate records across multiple files, where $\ome_j^2$, $a_\sig$, $b_\sig$, $\al_{\ell,m}$,  $\lambda$,  $a_\ell$, $b_\ell$ are all considered as known hyperparameters.



Given the characteristics of our examples in Section \ref{sec_motivating_examples}, we consider a specific class of linkage structures in which (a) records within the same file cannot be considered as corrupted versions from each other and (b) no more than two records correspond to the same latent individual. Such a class leads to partitions $\mathcal{C}_{\xiv}$ whose elements are either singletons or pairs (of records in different files). Having these restrictions in mind, we assume that every element of $\mathcal{C}_{\xiv}$ is equally likely a priori, which means that $\xiv$ is restricted to produce partitions composed of equally likely singletons and pairs in such a way that $p(\xiv)\propto 1$. Even though this assumption implies a non-uniform prior on related quantities, such as the number of individuals in the data, the uniform prior on $\xiv$ is convenient because it greatly simplifies computation of the posterior.


\subsection{Hyperparameter elicitation}\label{sec_hyperparameter_elicitation}

Careful elicitation of the hyperparameters is key to ensure appropriate model performance. For simplicity, we place a diffuse prior distribution on each $\beta_j$ by letting $\ome_j=100$. Then, following \citet[Section 2.4]{krivitsky-2008}, we set
$a_\sig$ and $b_\sig$ in such a way that a priori $\sig^2$ is vaguely concentrated (e.g., $\coefvar{\sig^2} = 0.5$) around $\expec{\sig^2}= \tfrac{\sqrt{I}}{\sqrt{I} - 2}\, \tfrac{\pi^{K/2}}{ \Gamma(K/2 + 1)}\, I^{2/K}$. Intuitively, each node needs a certain amount of space; since the volume of a $K$-dimensional hypersphere of radius $r$ is $\tfrac{\pi^{K/2}}{ \Gamma(K/2 + 1)}\, r^{K}$, it makes sense to set the prior standard deviation proportional to $I^{1/K}$.

As far as the profile hyperparameters is concerned, we set $\alv_{\ell}$ to either $\onev_{M_\ell}$, an $M_\ell$-dimensional vector of ones, or $\gamma_\ell$, the empirical frequencies of the corresponding values in field $\ell$, depending on whether or not field $\ell$ is taken as categorical independent of $\pi_{\xi_{i,j},\ell}$ or string-valued. Next, following \citet[Section 4]{steorts-2015-empirical}, we set $a_{\ell} = a = 1$ and $b_\ell = b = 99$, which corresponds to a prior mean of 0.01 for the distortion probabilities. This choice seems to be adequate since distortion probabilities cannot be close to 1 for a RL problem to be sensible (typically a small number of corrupted fields is expected), and based on our choice of the beta distribution for each $\psi_\ell$, it follows that $b \gg  1$. Finally, if string-valued fields are available, setting $\lambda = 1$ and letting $d(\cdot,\cdot)$ be the Edit distance \citep{christen-2012} in the distortion model \eqref{eq_emipirical_distr_string_field} is a common choice.


\subsection{Computation}

For a given value of $K$, Bayesian parameter estimation can be achieved via Markov chain Monte Carlo (MCMC) algorithms. The posterior distribution is approximated using dependent but approximately identically distributed samples $\UPS^{(1)}, \ldots, \UPS^{(S)}$ where
\begin{multline*}
\UPS^{(s)} = \left( \xiv_1^{(s)},\ldots,\xiv_J^{(s)},\uv_{1}^{(s)},\ldots,\uv_{N}^{(s)}, \beta_1^{(s)},\ldots,\beta_J^{(s)}, \sig^{(s)}, \right. \\
\left. \piv_1^{(s)}, \ldots,\piv_N^{(s)}, \vtev_1^{(s)},\ldots,\vtev_L^{(s)}, \wv_1^{(s)},\ldots,\wv_J^{(s)}, \psi_1^{(s)},\ldots,\psi_L^{(s)} \right)  .
\end{multline*}
Notice that we can make inferences about $\xiv$, or any function of it, by handling samples $\xiv^{(1)},\ldots,\xiv^{(S)}$ drawn from the posterior distribution $p(\UPS\mid\data)$. Details about posterior linkage estimation are given in Section \ref{sec_posterior_linkage}.


In order to derive the joint posterior we need to recall that $\pi_{n,\ell}$, $w_{i,j,\ell}$ and $\xi_{i,j}$ are all interconnected, since $w_{i,j,\ell} = 0$ implies that $\pi_{\xi_{i,j},\ell} = p_{i,j,\ell}$. However, if $\pi_{\xi_{i,j},\ell} = p_{i,j,\ell}$, then $w_{i,j,\ell}$ may or may not equal 0. Full conditional distributions are available in closed form for all the profile parameters. As far as the network parameters is concerned, random walk Metropolis-Hastings steps can be used. Details about the MCMC algorithm are discussed in Appendix \ref{ap_RL2}.


\subsection{Remarks}

Since estimating the posterior distribution of the linkage structure $\xiv$ is the main inference goal, lack of identifiability about the latent positions $\uv_n$ is not an issue here. On the other hand, notice that we are able to make inferences on any function of $\xiv$ through the posterior samples $\xiv^{(1)},\ldots,\xiv^{(S)}$. For example, the posterior mean of the population size $N$ is straightforward to calculate:
$$
\expec{N\mid\data} = \frac{1}{S}\sum_{s=1}^S \max\le\{\xiv^{(s)}\ri\}.
$$

Moreover, it is very easy to provide probabilistic statements about any pair of records. Recall that two records $(i_1,j_1)$ and $(i_2,j_2)$ match if and only if they point to the same latent individual (i.e., $\xi_{i_1,j_1} = \xi_{i_2,j_2}$). Hence, the posterior probability of a match can be computed as
$$
\pr{\xi_{i_1,j_1} = \xi_{i_2,j_2}\mid\data} = \frac{1}{S}\sum_{s=1}^S\ind{\xi_{i_1,j_1}^{(s)} = \xi_{i_2,j_2}^{(s)}}.
$$

As a final remark, the selection of the latent dimension $K$ can be carried out again employing information criteria methods such as the Deviance Information Criterion (DIC)  \citep{spiegelhalter2002bayesian,gelman-2014-information,spiegelhalter-2014} and the Watanabe-Akaike Information Criterion (WAIC) \citep{watanabe2010asymptotic,watanabe2013widely,gelman-2014-information}

\subsection{Posterior linkage and performance assessment}\label{sec_posterior_linkage}

Once posterior samples of $\xiv$ have been obtained, we need to provide a point estimate of the overall linkage structure, say $\xivhat$. From a decision-theoretic point of view, we need to define a loss function and then set $\xivhat$ to that configuration of $\xiv$ that minimizes the posterior expected loss.

We consider a couple of options to estimate the overall linkage structure. \citet[Section 4]{lau-2007} provide a general approach based on a pairwise coincidence loss function \citep{binder-1978,binder-1981} and binary integer programming. On the other hand \citet[Section 3.3]{steorts-2014-SMERED} develop a methodology based on squared error and absolute number of errors loss functions \citep{tancredi-2011} and what they refer to as most probable maximal matching sets.

We assess the performance of both methods using conventional precision and recall metrics. Given the ground truth about the linkage structure, there are four possible ways of how predictions about pairs of records can be classified: Correct links (true positives, TP), correct non-links (true negatives, TN), incorrect links (false positives, FP), and incorrect non-links (false negatives, FN). The usual definitions of recall and precision are:
$$
\text{Recall}=\frac{\TP}{\TP+\FN},\quad\text{Precision}=\frac{\TP}{\TP+\FP}.
$$

Since in practice most of the records are unique, the vast majority of record pairs are classified as TN in most practical RL settings. Hence, we aim to achieve the highest possible recall (true positive rate) while keeping precision (positive predictive value) close to the maximum. Alternatively, the two measures can be combined into a single metric, the $\text{F}_{1}$ score, given by
$$
\text{F}_1 = 2\cdot\frac{\text{Precision}\cdot\text{Recall}}{\text{Precision}+\text{Recall}},
$$
which is the harmonic mean of precision and recall. As is the case for recall and precision, the $\text{F}_{1}$ score reaches its best value at 1 and worst at 0.

%

\section{Re-identification example revisited}\label{sec_bartunov_dataset}


\begin{figure}[!t]
	\centering
	\subfigure[Seed node ID 11209.]{\includegraphics[scale=.27,]{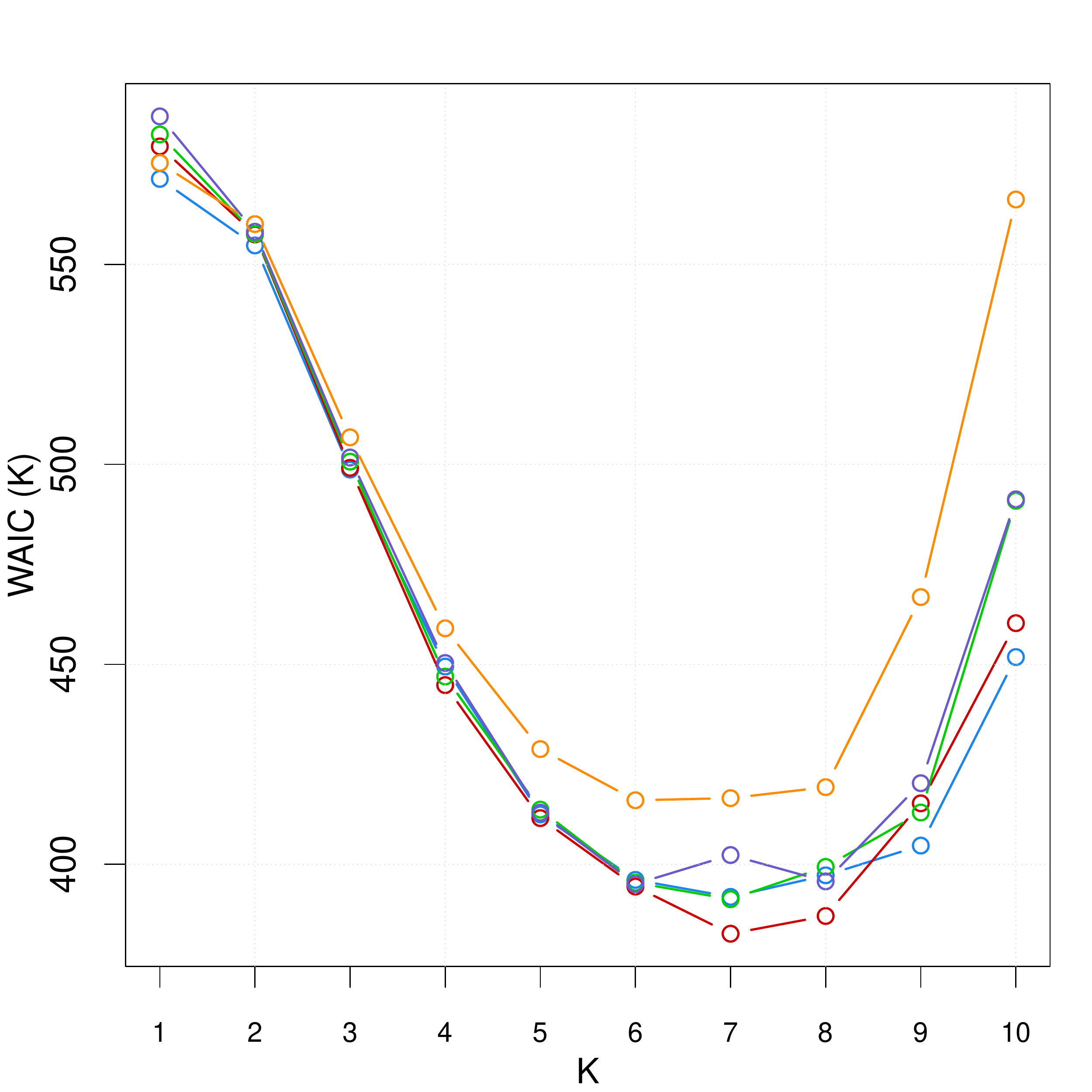}}
	\subfigure[Seed node ID 5099.]{\includegraphics[scale=.27,]{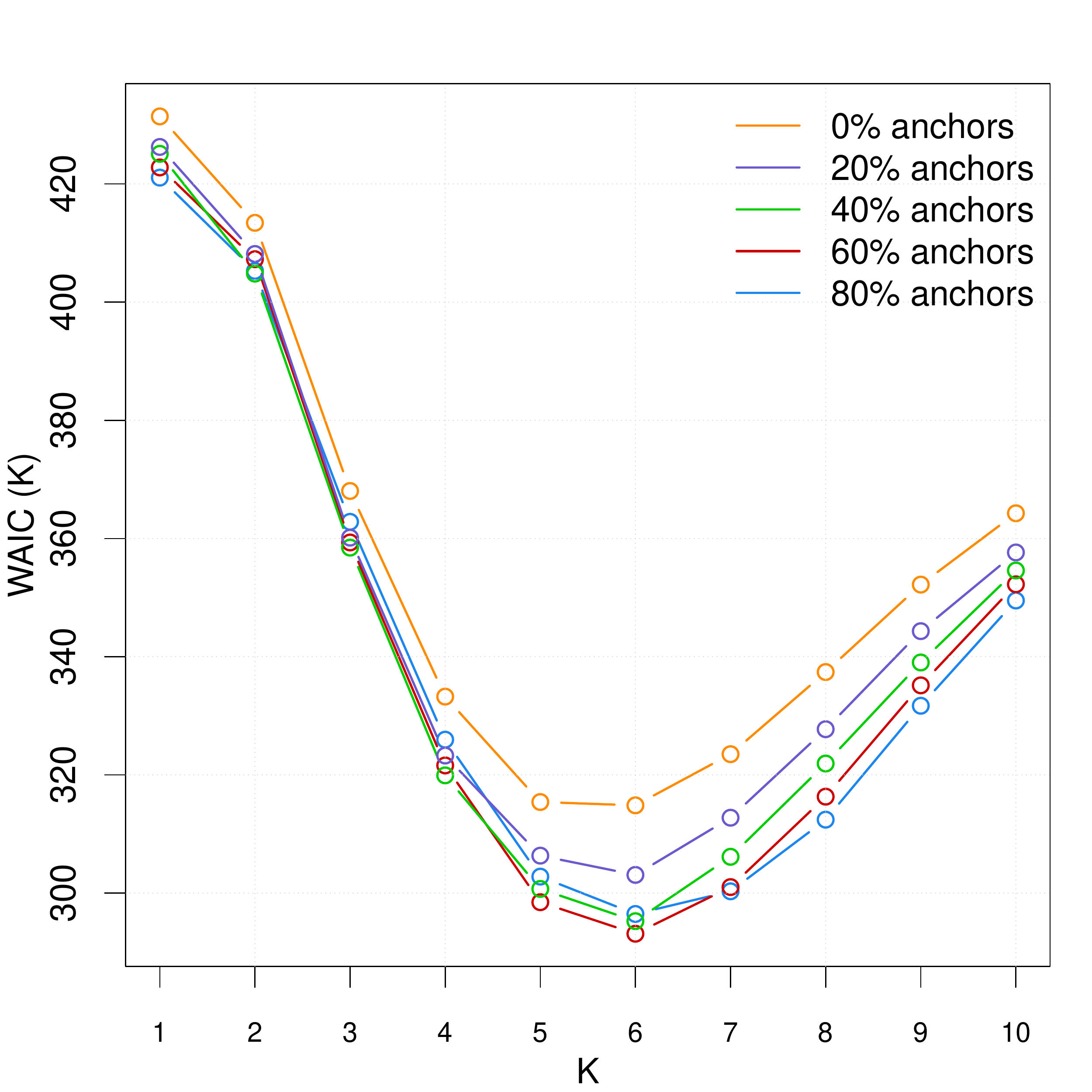}}
	\caption{\footnotesize{WAIC criteria for selecting the dimension $K$ of the latent space for two users in particular.}}
	\label{fig_WAIC_uir}
\end{figure}

In this Section we use the data introduced in Section \ref{sec_bartunov_dataset} to illustrate the ability of our model to match individuals across networks in the presence of network information alone.  We fit our model using a logit link and setting $K$ to that value favored by the information criteria discussed in Section \ref{sec_model_selection}.  Models are fitted assuming different numbers of anchor nodes, i.e., nodes that are known to match across both networks. All results shown below are based on 10,000 samples obtained after a thinning of 10 iterations and a burn-in period of 10,000 iterations. The clustering methodology proposed by \citet{lau-2007} was used to obtain a point estimate of the posterior linkage structure.


First, we analyze in detail two of the 17 pairs of networks. Figure \ref{fig_WAIC_uir} shows the values of the WAIC associated with models of different dimension and fraction of anchor nodes for two seed users (11209 and 5099). We use the optimum value of $K$ pointed out by this criterion every time we fit the model; for instance, a choice of $K=6$ is used for all of our analysis considering user 5099. It is reassuring that the dimension of the latent space is robust to the number of anchors, but it can clearly differ from one dataset to the other.

\begin{figure}[!h]
	\centering
	\includegraphics[scale=.27,]{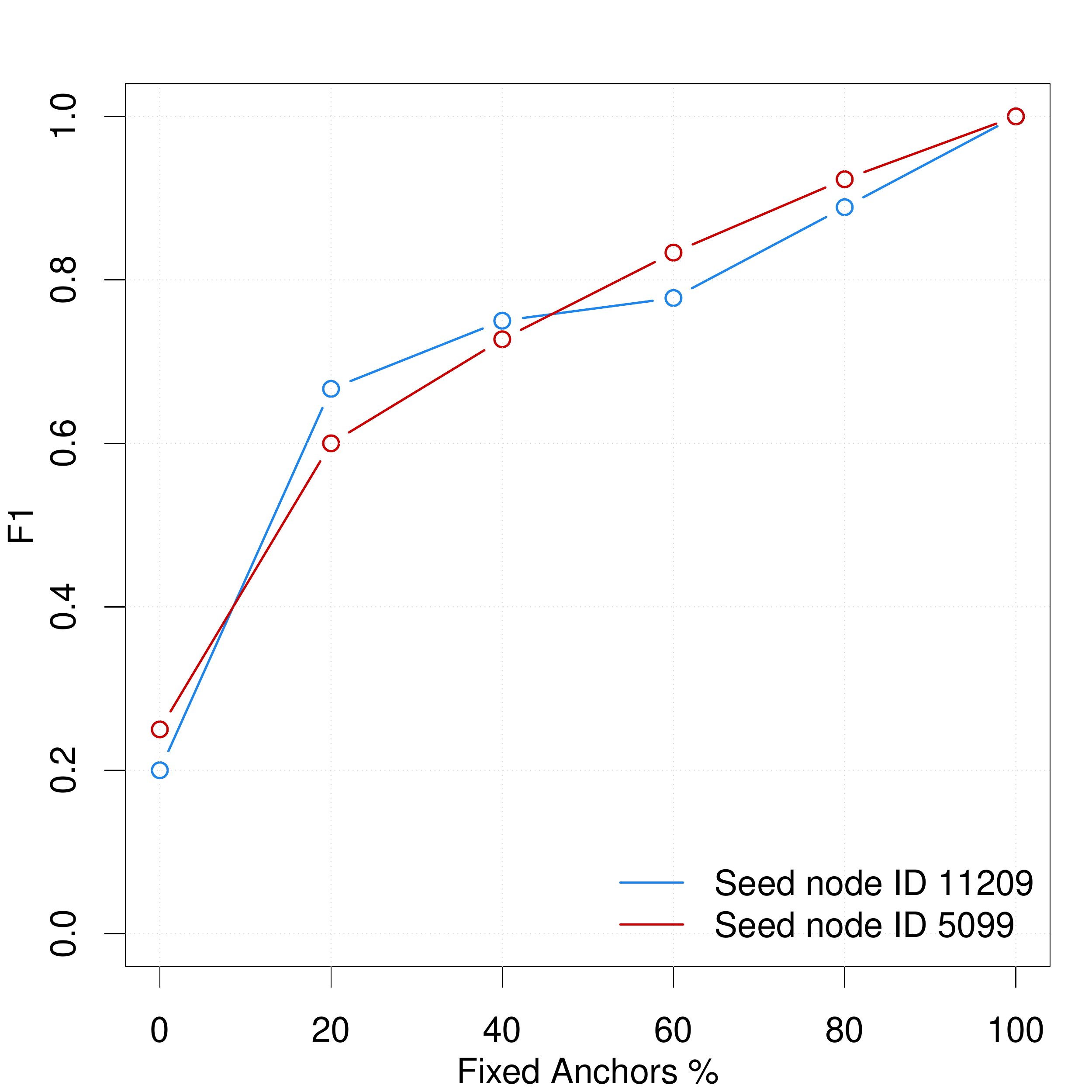}
	\caption{\footnotesize{Impact of anchor nodes fraction for our model for two users in particular. Profile information is missing. Values of $K$ selected according to those values favored by the WAIC in Figure \ref{fig_WAIC_uir}. }}
	\label{fig_anchors_plot_WAIC_uir}
\end{figure}

Figure \ref{fig_anchors_plot_WAIC_uir} shows plots of $\text{F}_1$ scores as a function of the anchor nodes fraction. Not surprisingly, Figure \ref{fig_anchors_plot_WAIC_uir} makes clear that it is very difficult to find correct matches just relying on link information with no anchors at all. However, the performance of the procedure improves considerably as more anchors are identified.

Next, we compare our latent model with the approach introduced in \citealp{bartunov-2012} (BARTUNOV for short) on the whole set of 17 pairs of networks. BARTUNOV is a non-symmetric supervised machine learning approach based on conditional random fields built on conditional random fields. Consider graphs $A = (V,E)$  where $V$ denotes the set of vertices (nodes) and $E$ denotes the set of edges. Let $\mu(v)$ be the projection of $v\in A$ to graph $B$. Under this approach observed variables are those nodes (profiles) to be projected in graph $A$, $\X=\{\x_v = v\mid v\in V\}$, while hidden variables are correct projections of these nodes, $\Y = \{\y_v=\mu(v)\mid v\in V\}$.  Here, the posterior probability of a projection is
$$
p(\Y\mid\X) \propto \ex{-\le(\sum_{v\in V} \Psi(\y_v\mid\x_v) + \sum_{(v,u)\in E} \Omega(\y_v,\y_u) \ri)},
$$
where $\Psi$ is an unary energy function (profile distance) and $\Omega$ a binary energy function (network distance). In anonymized settings with no profile attribute information, notice that $\Psi\equiv 0$ and only network information is used.

The binary energy function $\Omega$ ranges from 0 to 1 and represents network distance between projections of nodes $v$ and $u$:
$$
\Omega(\y_v,\y_u)=
\left\{
\begin{array}{ll}
\infty, & \hbox{if $\y_v=\y_u$;} \\
1 - \frac{2\,w(L_v\cap L_u)}{w(L_v) + w(L_u)}, & \hbox{otherwise,}
\end{array}
\right.
$$
where $L_v$ and $L_u$ are sets of neighbors of $v$ and $u$,
respectively, and the weighting function is defined as $w(L)=\sum_{v\in L} 1/\log(d(v))$, where $d(v)$ is the degree of node $v$. We refer the reader to Section 3 in \citet{bartunov-2012} for details about training classifiers and computation.

\begin{figure}[!h]
	\centering
	\subfigure[Recall.]{\includegraphics[scale=.36,]{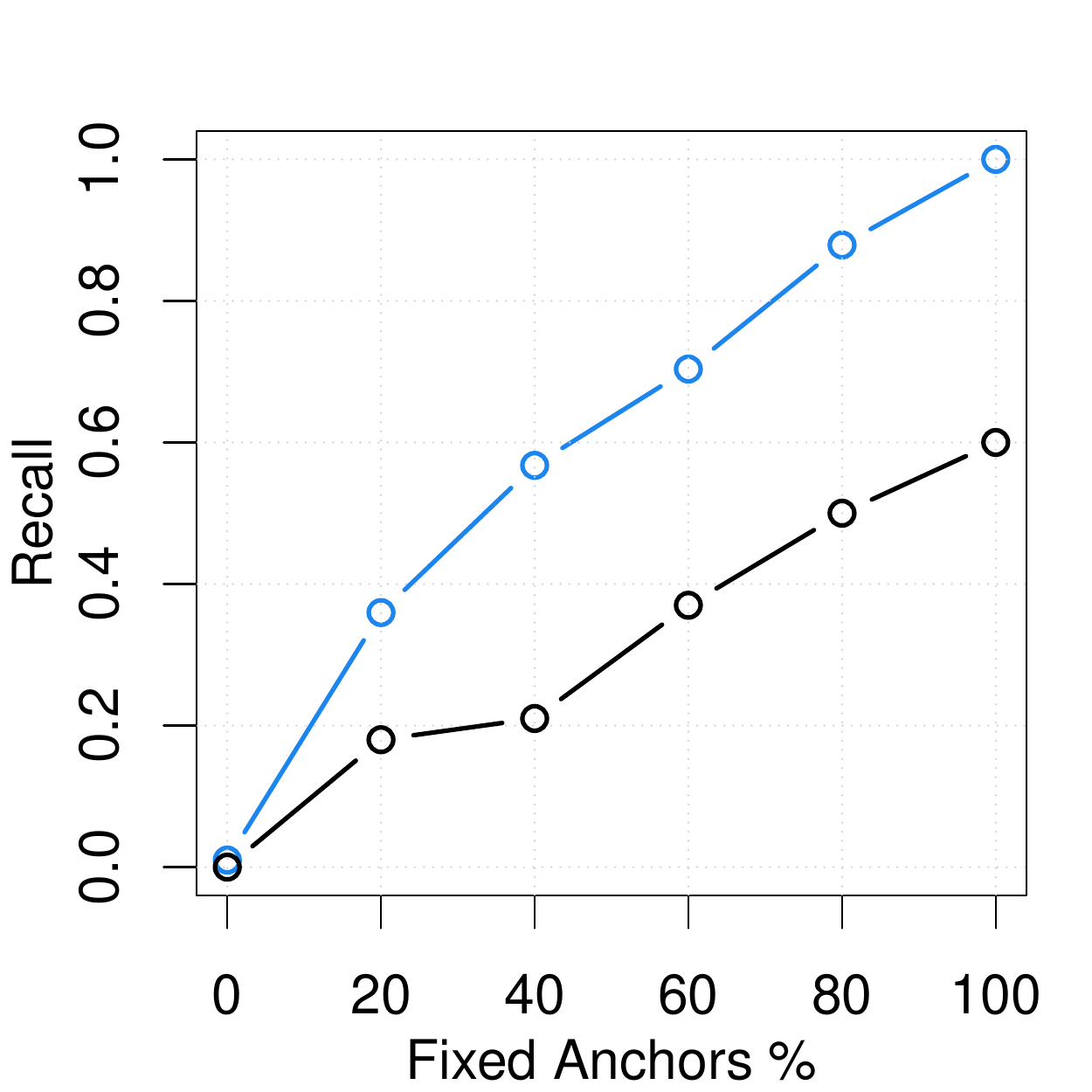}}
	\subfigure[Precision.]{\includegraphics[scale=.36,]{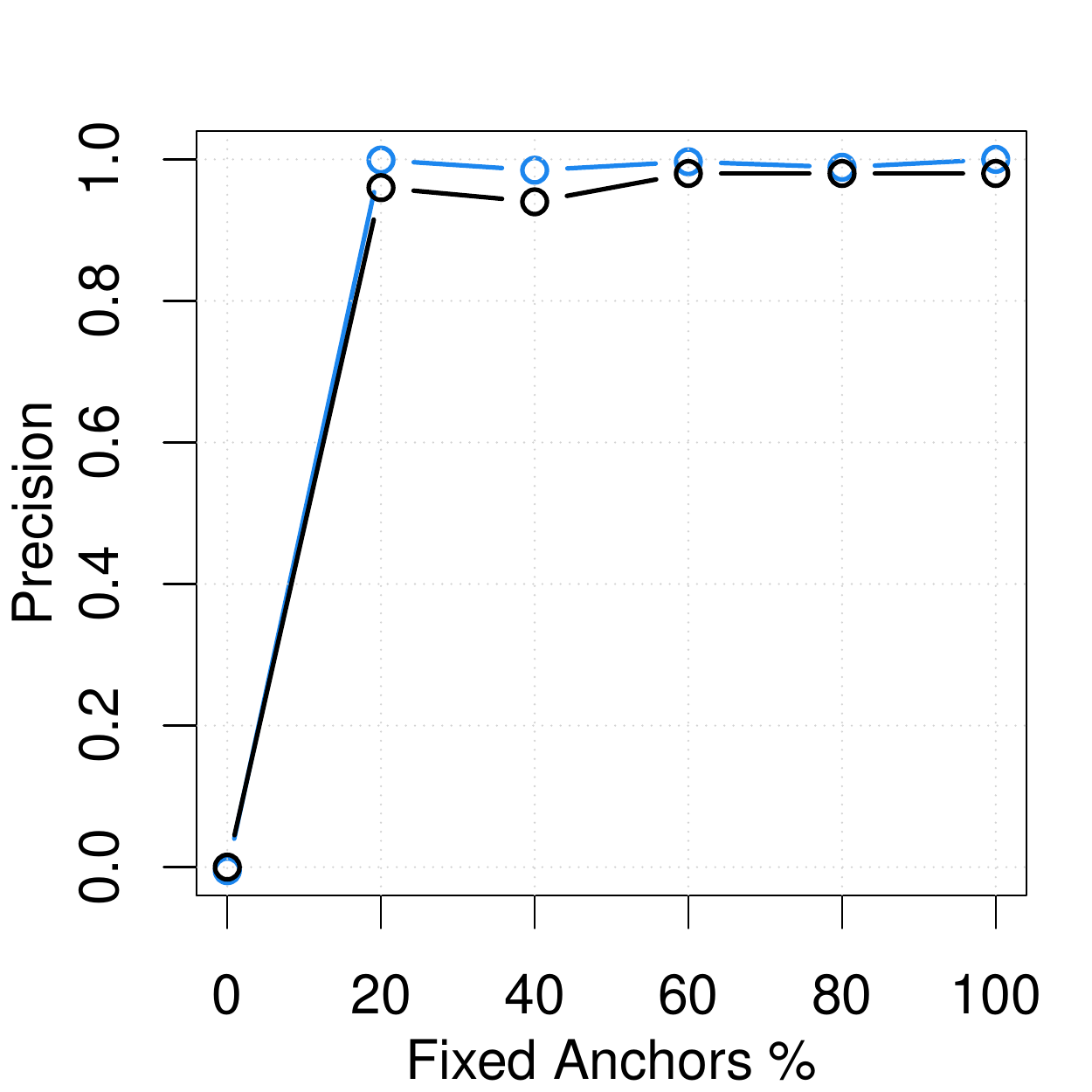}}
	\subfigure[$\text{F}_1$ score.]{\includegraphics[scale=.36,]{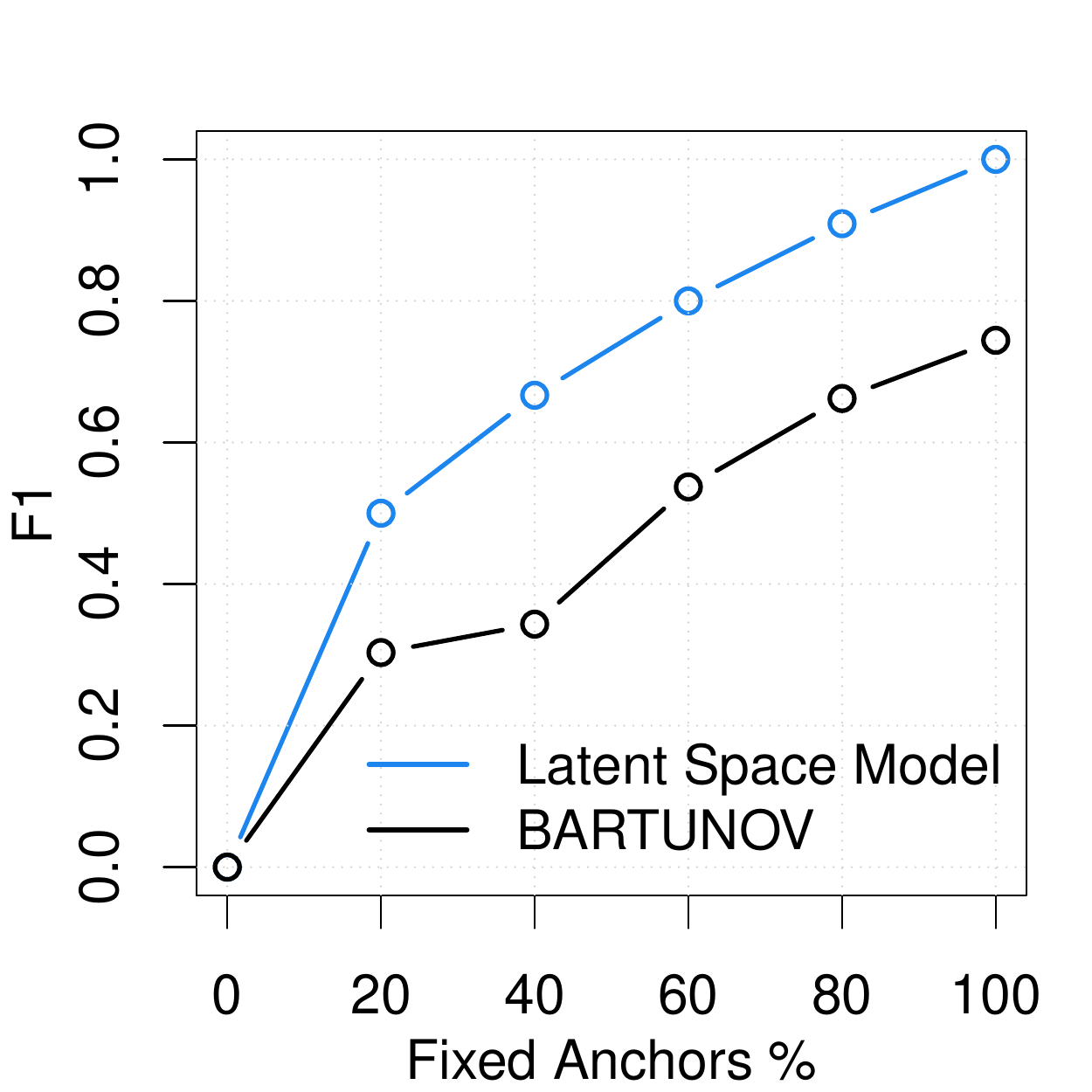}}
	\caption{\footnotesize{Impact of anchor nodes fraction for our latent space model and BARTUNOV. Profile information is missing.}}
	\label{fig_anchors_plot_uir}
\end{figure}

We fit both models to each pair of networks assuming different percentages of randomly selected anchors. For our model, we use $K=2$ in order to ease computation, but examine the robustness of the model later. Figure \ref{fig_anchors_plot_uir} shows the average precision, recall, and $\text{F}_1$ score over the 17 datasets. Neither our approach nor BARTUNOV are able to link records when no anchors are provided. When some anchors are fixed, the precision jumps and is practically the same for both alternatives, although our model seems to slightly outperform BARTUNOV. On the other hand, our approach returns a better performance in terms of recall, which means that the latent space model provides more correct links than its competitor. Lower recall values for BARTUNOV might be due to the fact that this approach is not symmetric and does not isolate the anchors from other records, in contrast to our model that automatically does so through the prior specification. Since our model tends to outperform BARTUNOV in terms of both recall and precision, it also does so in terms of $\text{F}_1$ score.


\begin{figure}[hbt!]
	\centering
	\includegraphics[scale=.27,]{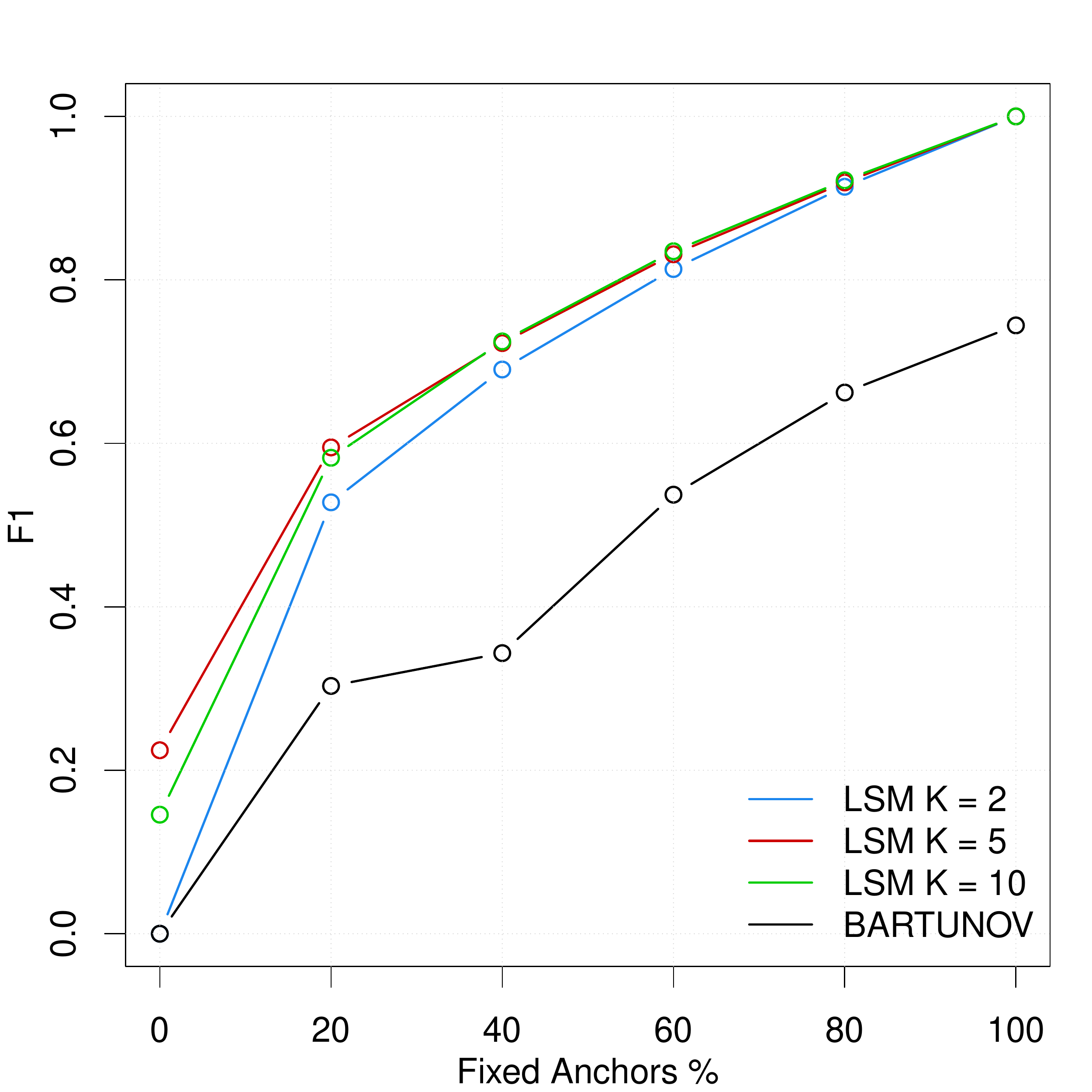}
	\caption{\footnotesize{Impact of anchor nodes fraction for our latent space model for $K=2,5,10$ and BARTUNOV. Profile information is missing.}}
	\label{fig_anchors_plot_sensitivity_uir}
\end{figure}

To assess the sensitivity of our results with respect to $K$, we fit our model again with both $K=5$ and $K=10$. Figure \ref{fig_anchors_plot_sensitivity_uir} shows a summary of the results for the $\text{F}_1$ score. As anticipated, we see that by fitting our model with higher dimensions, we slightly improve the accuracy of the posterior linkage, especially when the fraction of anchors is low; notice also that in all cases, our approach excels and outperforms BARTUNOV. These results suggest that, at least for this dataset, $K=5$ is a pragmatic choice that allow us to get satisfactory levels of accuracy, and including higher dimensions does not makes a substantial difference. Finally, note also that for both $K=5$ and $K=10$ our model is able to accurately classify some individuals even if no anchors are used.

\section{Identity resolution example revisited}\label{sec_buccafurri_dataset}

In this Section we analyze the data introduced in Section \ref{sec_buccafurri_dataset}. Recall that the main task in this setting is to identify as many accounts corresponding to the same user as possible. Here we illustrate how taking into account network information in such matching process can make a substantial difference. To this end, we fit our model using only profile data (profile model, PM), and also, using both profile and network data (profile and network model, PNM).

In this case, \verb"username" is the only profile information available; this attribute is considered as a string-valued field in \eqref{eq_model_RL_part_2}. As before, we use a logit link in \eqref{eq_model_RL_part_1} and set $K$ to that value favored by the information criteria discussed in Section \ref{sec_model_selection}. Results shown below are based on 10,000 samples obtained after a thinning of 10 iterations and a burn-in period of 10,000 iterations. The clustering methodology proposed by \citet{lau-2007} was used to obtain a point estimate of the posterior linkage structure.

Table \ref{tab_IC_buccafurri} presents the values of the DIC and WAIC associated with PNM for different dimensions of the latent space. Both criteria favor a choice of $K = 4$; we use such value for all of our analyzes in this Section.

\begin{table}[ht]
	\centering
	\begin{tabular}{c|ccccccc}
		\hline
		$K$  & 2 & 3 & 4 & 5 & 6 & 7 & 8 \\
		\hline
		DIC  & 12,805.9 & 12,473.5 & \textbf{12,415.5} & 12,960.9 & 13,911.2 & 14,067.66 & 14,172.0 \\
		WAIC & 13,784.2 & 13,168.5 & \textbf{13,113.7} & 13,322.1 & 13,945.6 & 14,080.24 & 14,177.4 \\
		\hline
	\end{tabular}
	\caption{\label{tab_IC_buccafurri}\footnotesize{ Values of DIC and WAIC for selecting the dimension $K$ of the latent space for our model in the identity resolution example.}}
\end{table}

Table \ref{tab_results_buccafurri} shows accuracy measures and summary statistics associated with both models. As expected, PNM exhibits higher values of recall and precision, which means that including network information in the analysis improves the accuracy of the posterior linkage. Notice also that PNM provides an estimate of the population size closer to the truth (585 in this case) and even more precise.
These findings strongly suggest that combining both sources of information at the cost of incrementing the parameter space is in fact helpful and worth the extra effort.

\begin{table}[!h]
	\centering
	\begin{tabular}{c|ccccc}
		\hline
		Model & Recall & Precision & $\text{F}_1$ & $\expec{N\mid\text{data}}$ & $\sd{N\mid\text{data}}$ \\
		\hline
		PM    &  0.26 & 0.62 & 0.37 & 599.67 & 7.94 \\
		PNM   &  0.52 & 0.71 & 0.60 & 592.06 & 5.74 \\
		\hline
	\end{tabular}
	\caption{\label{tab_results_buccafurri}\footnotesize{Performance assessment and summary statistics for our model in the identity resolution example. Models using only profile data (profile model, PM), and also, using both profile and network data (profile and network model, PNM) are considered.}}
\end{table}


\section{Discussion}\label{sec_extensions}

We have introduced a novel approach for linking multiple files using both profile and network information. In particular, we have developed relevant aspects of the model including sampling algorithms and point estimates of the overall linkage structure. Our findings suggest that our methodology can produce accurate point estimates of the linkage structure even in the absence of profile information.


Our model can be extended in several ways. We could consider distortion probabilities of the form $\psi_{\ell,j}$ depending on both files and fields instead of just fields as in $\psi_\ell$; this modification reflects the fact that different files may be more or less prone to error. Alternatively, in an extra level of refinement, we could also weaken the independence assumption among fields by adding an extra layer of structure to account for dependencies; although as a result the parameter space and the complexity of the model would increase substantially. In addition, we could devise more general priors over the linkage structure that allow us to generate clusters whose sizes grow sublinearly with the size of the data, which is typically the case in RL applications, and also handle RL and de-duplication problems simultaneously. \citet{miller-2015} and \citet{betancourt-2016} have already introduced two alternatives in order to address such micro-clustering requirement.

Finally, even though Bayesian methods consider flexible generative models that share power across databases, they mainly rely on MCMC methods, which could potentially turn out to be too slow on big databases. For instance, \citet{zhang-2015} share a dataset with more than 44 million users; the main task is to find as many duplicate users as possible across OSNs as in our identity resolution illustration. Such a dataset makes clear that it is worth considering fast approximation techniques either in the flavor of variational approximations (\citealp{saul-1996}, \citealp{jordan-1998}, \citealp{beal-2003}, \citealp{broderick-2014}) or stochastic gradient dynamics (\citealp{welling-2011}, \citealp{ahn-2012}, \citealp{chen-2014}).

\bibliographystyle{apalike}
\bibliography{references}

\section*{Computation}\label{ap_RL2}

For simplicity, we consider first the case in which every field is taken as a categorical field. Taking this into account that $\pi_{n,\ell}$, $w_{i,j,\ell}$ and $\xi_{i,j}$ are all interconnected, since if $w_{i,j,\ell} = 0$, then it must be the case that $\pi_{\xi_{i,j},\ell} = p_{i,j,\ell}$, the joint posterior reduces to:
\begin{equation*}
{\small
	\begin{aligned}
	p(\UPS\mid\Y,\P)
	&\propto
	\prod_{j,i<i'} \theta_{i,i',j}^{y_{i,i',j}}(1-\theta_{i,i',j})^{1-y_{i,i',j}}\times \prod_j \ex{-\tfrac{1}{\ome_j^2}\,\beta_j^2}\\
	&\,\,\,\times \prod_n (\sig^2)^{-K/2}\,\ex{-\tfrac{1}{\sig^2}\,\|\uv_n\|^2}\times (\sig^2)^{-(a_\sig+1)}\,\ex{-\frac{b_\sig}{\sig^2}}\\
	&\,\,\,\times\prod_{j,i,\ell,m} \le[ (1-w_{i,j,\ell})\, \del_{\pi_{\xi_{i,j},\ell}}(p_{i,j,\ell}) + w_{i,j,\ell}\,\vartheta_{\ell,m}^{\ind{p_{i,j,\ell} = m^{(i,j)}}} \ri]\\
	&\,\,\,\times \prod_{j,i,\ell} \psi_\ell^{w_{i,j,\ell}}(1-\psi_\ell)^{1-w_{i, j, \ell}}\times \prod_{n,\ell,m} \vartheta_{\ell,m}^{\ind{\pi_{n,\ell} = m^{(i,j)}}}\\
	&\,\,\,\times \prod_{\ell,m} \vartheta_{\ell,m}^{\al_{\ell,m} - 1} \times \prod_{\ell} \psi_\ell^{a_\ell - 1}(1-\psi_\ell)^{b_\ell - 1}  \times p(\xiv)
	\end{aligned}
}
\end{equation*}
where $\theta_{i,i',j} = g^{-1}(\be_j - \|\uv_{\xi_{i,j}} - \uv_{\xi_{i',j}}\|)$ and $m^{(i,j)}$ is the (category) index in which $p_{i,j,\ell}$ is having a value.

Our MCMC algorithm iterates over the model parameters $\UPS$. Where possible we sample from the full conditional posterior distributions as in Gibbs sampling; otherwise we use Metropolis-Hastings steps. The MCMC algorithm proceeds by generating a new state $\UPS^{(s+1)}$ from a current state $\UPS^{(s)}$ as follows:
\begin{enumerate}
	\item Sample $\xi_{i,j}^{(s+1)}$ following Algorithm 5 given in \citet{neal-2000}: Repeat the following update of $\xi_{i,j}$ $R$ times:
	\begin{enumerate}[i.]
		\item Draw a proposal, $\xi_{i,j}^{*}$, uniformly at random from the set of values or following a (fixed) distribution independent of the current state of the Markov chain; however, crucially, it may depend on the observed data. The proposal $\xi_{i,j}^{*}$ must leave the members of $\mathcal{C}_{\xiv}$ only having one (singletons) or two (pairs) records from different files.
		\item If $\xi_{i,j}^{*}$ stars a new cluster, sample a value for $\pi_{\xi_{i,j}^{*},\ell}$ from $\Cat(\vtev_\ell)$ for each $\ell$, and equivalently, sample a value for $u_{\xi_{i,j}^{*},k}$ from $\Nor(0,\sig^2)$ for each $k$.
		\item Compute the acceptance probability
		$$
		a= \min\left[1, \frac{p(\Y,\uv_{\xi_{i,j}^{*}}\mid\rest)}{p(\Y,\uv_{\xi_{i,j}^{(s)}}\mid\rest)} \right].
		$$
		\item Let
		$$
		\xi_{i,j}^{(s+1)} =
		\left\{
		\begin{array}{ll}
		\xi_{i,j}^{*}, & \hbox{with probability $a$;} \\
		\xi_{i,j}^{(s)}, & \hbox{with probability $1-a$.}
		\end{array}
		\right.
		$$
		\item If $\xi_{i,j}^{(s+1)} = \xi_{i,j}^{*}$, then $w_{i,j,\ell}$ has to be sampled accordingly; the same way that $\pi_{\xi_{i,j}^{*},\ell}$ and  $u_{\xi_{i,j}^{*},k}$ need to be updated with the values drawn from the prior in ii.
	\end{enumerate}
	This step depends on the form of $p(\xiv)$ and may involve additional parameters that also need to be sampled. See details below.
	
	\item Sample $w_{i, j, \ell}^{(s+1)}$ from $p(w_{i, j, \ell}\mid\rest) = \Ber(w_{i, j, \ell}\mid\tilde{\theta}_{i,j,\ell})$ where
	$$
	\tilde{\theta}_{i,j,\ell} =
	\left\{
	\begin{array}{ll}
	1, & \hbox{if $p_{i, j, \ell} \neq \pi_{\xi_{i,j},\ell} $;} \\
	\frac{\psi_\ell\displaystyle\prod_m\vte_{\ell,m}^{\ind{p_{i, j, \ell} = m^{(i,j)} }}}{\psi_\ell\displaystyle\prod_m\vte_{\ell,m}^{\ind{p_{i, j, \ell} = m^{(i,j)} }} + (1-\psi_\ell)}, & \hbox{if $p_{i, j, \ell} = \pi_{\xi_{i,j},\ell} $.}
	\end{array}
	\right.
	$$
	\item Sample $\pi_{n,\ell}^{(s+1)}$ from $p(\pi_{n,\ell}\mid\rest) = \del_{p_{i, j, \ell}}(\pi_{n,\ell}\mid p_{i, j, \ell})$ if there exists a file $j$ and a record $i\in C_{n,j}$ such that $w_{i, j, \ell} = 0$, where $C_{n,j} = \{i\in [I_j]:\xi_{i,j} = n\}$, i.e., $C_{n,j}$ is the set of all records $i$ in file $j$ linked to the latent individual $n$. Otherwise, sample $\pi_{n,\ell}^{(s+1)}$ from $p(\pi_{n,\ell}\mid\rest) = \Cat(\pi_{n,\ell}\mid\vtev_\ell)$.
	\item Sample $\uv_n^{(s+1)}$:
	\begin{enumerate}[i.]
		\item Draw a proposal, $\uv_n^*$, from $\Nor(\uv_n^{(s)},\delta^2\I_K)$ where $\delta^2$ is a tunning parameter.
		\item Compute the acceptance probability
		$$
		a =
		\min\left[1, \frac{p(\Y,\uv_n^{*}\mid\rest)}{p(\Y,\uv_n^{(s)}\mid\rest)} \right].
		$$
		\item Let
		$$
		\uv_n^{(s)} =
		\left\{
		\begin{array}{ll}
		\uv_n^{*}, & \hbox{with probability $a$;} \\
		\uv_n^{(s)}, & \hbox{with probability $1-a$.}
		\end{array}
		\right.
		$$
	\end{enumerate}
	\item Sample $\psi_\ell^{(s+1)}$ from
	$$
	p(\psi_\ell\mid\rest) = \Bet\left(\psi_\ell\,\,\Big|\,\,a_\ell + \sum_{j,i}w_{i, j, \ell}, b_\ell + I - \sum_{j,i}w_{i, j, \ell}\right)
	$$
	where $I = \sum_j I_j$.
	\item Sample $\vtev_\ell^{(s+1)}$ from $p(\vtev_\ell\mid\rest) = \Dir(\vtev_\ell\mid\tilde{\alv})$ where
	$$
	\tilde{\al}_m = \al_{\ell,m} + \sum_n\ind{\pi_{n,\ell} = m} + \sum_{j,i}w_{i, j, \ell}\,\ind{p_{i, j, \ell} = m}
	$$
	where $\tilde{\alv}=[\tilde{\al}_1,\ldots,\tilde{\al}_{M_\ell}]^T.$
	\item Sample $\be_j^{(s+1)}$:
	\begin{enumerate}[i.]
		\item Draw a proposal, $\beta_j^*$, from $\Nor(\beta_j^{(s)},\delta^2)$ where $\delta^2$ is a tunning parameter.
		\item Compute the acceptance probability
		$$
		a =
		\min\left[1, \frac{p(\Y,\beta_j^*\mid\rest)}{p(\Y,\beta_j^{(s)}\mid\rest)} \right].
		$$
		\item Let
		$$
		\beta^{(s+1)} =
		\left\{
		\begin{array}{ll}
		\beta_j^*, & \hbox{with probability $a$;} \\
		\beta_j^{(s)}, & \hbox{with probability $1-a$.}
		\end{array}
		\right.
		$$
	\end{enumerate}
	\item Sample $(\si^2)^{(s)}$ from
	$p(\sig^2\mid\rest) = \IGamd\le(\sig^2 \mid a_\sig + \tfrac{N\,K}2, b_\sig + \tfrac12 \textstyle\sum_n\|\uv_n\|^2 \ri).$
\end{enumerate}

The value of the tunning parameter $\delta^2$ in the proposal distribution is chosen to make the algorithm run efficiently. We adaptively change the value of $\delta^2$ at the beginning of the chain in order to automatically find a good proposal distribution. See \citet{rosenthal-2011} for a review about optimal proposal scalings for Metropolis-Hastings MCMC algorithms and adaptive MCMC algorithms.

On the other hand, if field $\ell$ is a string-based field, then the following steps should be considered:
\begin{enumerate}
	\item If $p_{i, j, \ell} \neq \pi_{\xi_{i,j},\ell}$, then $w_{i, j, \ell}^{(s+1)} = 1$. If instead $p_{i, j, \ell} = \pi_{\xi_{i,j},\ell}$, sample $w_{i, j, \ell}^{(s+1)}$ from $p(w_{i, j, \ell}\mid\rest) = \Ber\le(w_{i, j, \ell}\mid q_{i,j,\ell}/(q_{i,j,\ell} + 1 - \psi_\ell)\ri)$ with
	$$q_{i,j,\ell} = \psi_\ell\,\gamma_\ell(p_{i, j, \ell})\,h_\ell(\pi_{\xi_{i,j},\ell})\,\ex{-\lambda\,d(p_{i, j, \ell},\pi_{\xi_{i,j},\ell})},$$
	where $h_\ell(s_0)=\le(\sum_s\ex{-\lambda\,d(s,s_0)}\ri)^{-1}$.
	
	\item  Sample $\pi_{n,\ell}^{(s+1)}$ from $p(\pi_{n,\ell}\mid\rest) = \del_{p_{i, j, \ell}}(\pi_{n,\ell}\mid p_{i, j, \ell})$ if there exists a file $j$ and a record $i$ such that $(i,j)\in R_{n}$ such that $w_{i, j, \ell} = 0$, where $R_{n} = \{(i,j):\xi_{i,j} = n\}$, i.e., $R_{n}$ is the set of all records $i$ in file $j$ linked to the latent individual $n$. Otherwise, sample $\pi_{n,\ell}^{(s+1)}$ from $p(\pi_{n,\ell}\mid\rest)$ where
	$$
	p(\pi_{n,\ell} = s \mid\rest) \propto \ga_\ell(s)\,\ex{\sum_{R_n}w_{i,j,\ell}\,\le( \log h_\ell(s) - \lambda\,d(p_{i,j,
			\ell},s) \ri)}.
	$$
\end{enumerate}

\end{document}